\documentclass[amsmath,amssymb,reprint,aps,prl,floatfix,superscriptaddress]{revtex4-1}
\usepackage{bm}
\usepackage{amsmath}
\usepackage{graphicx}
\graphicspath{{./sm/} }
\usepackage{braket}
\usepackage{epstopdf}
\usepackage{bbm}
\usepackage{hyperref}

\hypersetup{
    colorlinks,
    linkcolor={black},
    citecolor={black},
    urlcolor={black},
    pdfborder={0 0 0}
    }

\begin{document}

\title{Quantum-enhanced interferometry with large heralded photon-number states}

\author{G.S. Thekkadath}
\affiliation{Clarendon Laboratory, University of Oxford, Parks Road, Oxford OX1 3PU, UK}
\email{guillaume.thekkadath@physics.ox.ac.uk}

\author{M.E. Mycroft}
\affiliation{Faculty of Physics, University of Warsaw, ul. Pasteura 5, 02-093 Warsaw, Poland}

\author{B.A. Bell}
\affiliation{Clarendon Laboratory, University of Oxford, Parks Road, Oxford OX1 3PU, UK}

\author{C.G. Wade}
\affiliation{Clarendon Laboratory, University of Oxford, Parks Road, Oxford OX1 3PU, UK}

\author{A. Eckstein}
\affiliation{Clarendon Laboratory, University of Oxford, Parks Road, Oxford OX1 3PU, UK}

\author{D.S. Phillips}
\affiliation{Clarendon Laboratory, University of Oxford, Parks Road, Oxford OX1 3PU, UK}

\author{R.B. Patel}
\affiliation{Clarendon Laboratory, University of Oxford, Parks Road, Oxford OX1 3PU, UK}

\author{A. Buraczewski}
\affiliation{Faculty of Physics, University of Warsaw, ul. Pasteura 5, 02-093 Warsaw, Poland}

\author{A.E. Lita}
\affiliation{National Institute of Standards and Technology, 325 Broadway, Boulder, Colorado 80305, USA}

\author{T. Gerrits}
\affiliation{National Institute of Standards and Technology, 325 Broadway, Boulder, Colorado 80305, USA}
\affiliation{National Institute of Standards and Technology, 100 Bureau Drive, Gaithersburg, Maryland 20899, USA}

\author{S.W. Nam}
\affiliation{National Institute of Standards and Technology, 325 Broadway, Boulder, Colorado 80305, USA}

\author{M. Stobi\'{n}ska}
\affiliation{Faculty of Physics, University of Warsaw, ul. Pasteura 5, 02-093 Warsaw, Poland}

\author{A.I. Lvovsky}
\affiliation{Clarendon Laboratory, University of Oxford, Parks Road, Oxford OX1 3PU, UK}

\author{I.A. Walmsley}
\affiliation{Clarendon Laboratory, University of Oxford, Parks Road, Oxford OX1 3PU, UK}
\affiliation{Department of Physics, Imperial College London, Prince Consort Rd, London SW7 2AZ, UK}

\begin{abstract}
Quantum phenomena such as entanglement can improve fundamental limits on the sensitivity of a measurement probe.
In optical interferometry, a probe consisting of $N$ entangled photons provides up to a $\sqrt{N}$ enhancement in phase sensitivity compared to a classical probe of the same energy.
Here, we employ high-gain parametric down-conversion sources and photon-number-resolving detectors to perform interferometry with heralded quantum probes of sizes up to $N=8$ (i.e. measuring up to 16-photon coincidences).
Our probes are created by injecting heralded photon-number states into an interferometer, and in principle provide quantum-enhanced phase sensitivity even in the presence of significant optical loss.
Our work paves the way towards quantum-enhanced interferometry using large entangled photonic states.
\end{abstract}

\maketitle

\section*{Introduction}

Optical interferometry provides a means to sense very small changes in the path of a light beam. 
These changes may be induced by a wide range of phenomena, from pressure and temperature variations that impact refractive index, to modifications of the space-time metric that characterize gravitational waves.
In its simplest form, interferometry measures distortions via the phase difference $\phi$ between the two paths of the interferometer.
The uncertainty $\Delta \phi$ in a measurement of this phase difference is limited fundamentally by the quantum noise of the illuminating light beams.
This noise can be reduced by employing light exhibiting nonclassical properties such as entanglement and squeezing in order to improve the sensitivity of an interferometer beyond classical limits~\cite{RDD2015review}.
Quantum states of light are most effective when it is desirable to maximize the phase sensitivity per photon inside an interferometer, such as in gravitational wave detectors~\cite{caves1981quantum,tse2019quantum} or when characterizing delicate photosensitive samples~\cite{crespi2012measuring,wolfgramm2013entanglement,taylor2013biological,taylor2016quantum,cimini2019adaptive}.

In principle, $N$-photon quantum states of light such as the highly entangled N00N state can provide up to a $\sqrt{N}$ precision enhancement over a classical state of equal energy~\cite{bollinger1996noon,mitchell2004super,walther2004broglie,nagata2007beating,kim2009three,afek2010high,matthews2011heralding,ulanov2016loss,slussarenko2017unconditional}.
Unfortunately these highly entangled states are vulnerable to decoherence, especially at large photon numbers.
In practice, their enhanced sensitivity disappears in the presence of loss which may originate from interactions inside the interferometer (e.g. absorption in a sample) as well as external losses in the state preparation and detection~\cite{datta2011quantum}.

Although a $\sqrt{N}$ enhancement is not achievable in the presence of loss, one can engineer states that trade-away sensitivity for loss-tolerance in order to still achieve some advantage over classical limits~\cite{dorner2009optimal,kacprowicz2010experimental}.
For example, squeezed light~\cite{demkowicz2013fundamental,lang2013optimal,yonezawa2012quantum,berni2015ab} and non-maximally entangled states such as Holland-Burnett states~\cite{holland1993interferometric,sun2008experimental,xiang2011entanglement,thomas2011realworld,xiang2013optimal,jin2016detection,matthews2016towards} can surpass classical limits despite some losses.
Importantly, the precision enhancement achievable with such states can grow with $N$, even in the presence of loss~\cite{dorner2009optimal}. 
%Mention that all below are Holland-Burnett states?
Experimental demonstrations have prepared unheralded $N=6$~\cite{xiang2013optimal,jin2016detection} (or heralded $N=2$~\cite{thomas2011realworld}) Holland-Burnett states, but further increase of $N$ is constrained by source brightness as well as detector efficiency and number-resolution.
This motivates developing experimental protocols that can produce and detect loss-tolerant states with larger photon numbers.

In this work, we address a number of key challenges in order to scale-up quantum-enhanced interferometry using definite photon-number states of light.
Firstly, we introduce probe states that are prepared by combining two photon-number states on a beam splitter similarly to Holland-Burnett states.
However, unlike the latter, we allow the initial photon-number states to be unequal.
We show that these generalized Holland-Burnett states are more sensitive than both Holland-Burnett and N00N states in the presence of loss and approximate the performance of the optimal probe~\cite{dorner2009optimal}.
Secondly, we experimentally implement our scheme using high-gain parametric down-conversion sources~\cite{eckstein2011highly,harder2016mesoscopic} and state-of-the-art photon-number-resolving detectors~\cite{lita2008counting} in order to access a large photon-number regime.
We herald entangled probes of sizes up to $N=8$ and measure up to 16-photon coincidences, thereby further increasing the scale of experimental multiphoton quantum technologies~\cite{gao2010experimental,wang2016experimental,wang2019boson}.

\begin{figure}
    \centering
    \includegraphics[width=1\columnwidth]{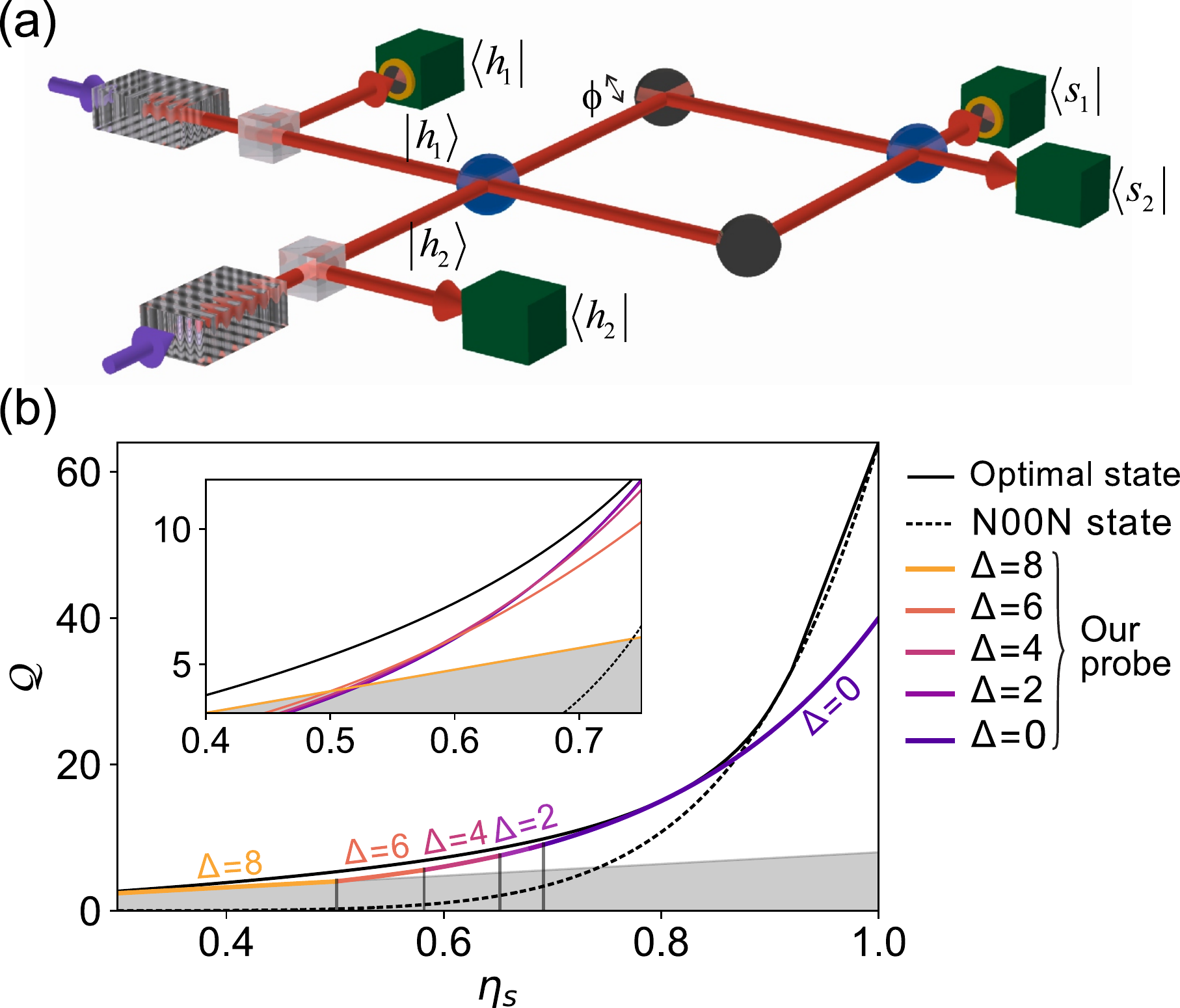}
	\caption{
    \textbf{Interferometric scheme}. (a) Two type-II parametric down-conversion sources each produce orthogonally-polarized pairs of beams that are separated using polarizing beam splitters.
    By measuring one of the beams from each source with a photon-number-resolving detector, we herald a pair of photon-number states $\ket{h_1,h_2}$.
    We inject this probe into an interferometer and perform photon counting at the output to estimate the unknown phase difference $\phi$.
    (b) Quantum Fisher information $\mathcal{Q}$ calculated for 8-photon ($N=h_1+h_2=8$) probes inside the interferometer as a function of the signal transmissivity $\eta_s$ which is assumed to be equal in both interferometer modes. 
    Coloured curve in the main figure plots $\mathcal{Q}$ of the probe with the optimal $\Delta = |h_1-h_2|$ for a given $\eta_s$, while the inset shows the full curves of each probe for $\eta_s \in [0.4,0.75]$.
    Our probe approximates the performance of the optimal state [black line] and surpasses that of the N00N state [dashed line] for efficiencies below $\sim 90 \%$.
    The grey filled region indicates performance below the shot-noise limit.
	}
	\label{fig:Scheme}
\end{figure}

The idea is illustrated in Fig.~\ref{fig:Scheme}(a).
Two type-II parametric down-conversion (PDC) sources each produce pairs of light beams that are quantum-correlated in photon number, i.e. a two-mode squeezed vacuum state
\begin{equation}
    \ket{\chi} = \sqrt{1-\lambda^2}\sum_{n=0}^\infty \lambda^n \ket{n,n}.
    \label{eq:tmsv}
\end{equation}
Here, $\lambda$ is a parameter that determines the average number of photons in each beam, $\braket{n} = \lambda^2/ (1-\lambda^2)$.
Measuring one of the beams with an ideal lossless photon-number-resolving detector projects the second beam onto a known photon-number state $\ket{h_1}$.
Duplicating this procedure with a second independent source and detector, we herald pairs of photon-number states that are not necessarily identical, i.e. the probe $\ket{h_1,h_2}$. 
When these states are combined on the first beam splitter, multiphoton interference generates a path-entangled probe inside the interferometer~\cite{stobinska2019quantum}.

We quantify the phase-sensitivity of the probe inside the interferometer by calculating the quantum Fisher information $\mathcal{Q}$. The quantity $\mathcal{Q}$ provides a lower limit on the best achievable phase uncertainty via the quantum Cramer-Rao bound, $\Delta \phi \geq 1/\sqrt{\mathcal{Q}}$.
The bound can be saturated using the optimal measurement strategy, which in the absence of loss is photon counting for the probes considered here~\cite{hofmann2009all,zhong2017optimal}. 

In Fig.~\ref{fig:Scheme}(b), we plot $\mathcal{Q}$ for several probes with the same total photon number $N=h_1+h_2=8$, but different $\Delta= |h_1 - h_2|$, as a function of the signal transmissivity $\eta_s$ which we assume to be equal in both interferometer modes.
Probes with a small $\Delta$ provide a greater advantage over the classical shot-noise limit but are more sensitive to losses.
Since the probe is heralded in our scheme, one can choose the optimal $\Delta$ for a given $\eta_s$. 

Also shown in  Fig.~\ref{fig:Scheme}(b) is $\mathcal{Q}$ for the optimal state that maximizes this parameter for a given $N$ and $\eta_s$. 
This state has been found in Ref.~\cite{dorner2009optimal}; the derivation is reproduced in the Supplementary Method 1. 
For the loss-free case ($\eta_s=1$), the optimal state is the N00N state. 
However, for efficiencies below $\sim 90 \%$, our probes significantly surpass the N00N state in terms of $\mathcal{Q}$, exhibiting performance close to optimal. 
Moreover, in contrast to the N00N and Holland-Burnett states, our probe performs at least as well as the shot-noise limit for any amount of loss.

We now turn to the experiment.
Both PDC sources are periodically poled potassium titanyl phosphate (ppKTP) waveguides pumped with $\sim 0.5$ ps long pulses from a mode-locked laser at a repetition rate of 100 kHz.
The four detectors are superconducting transition edge sensors which we use to count up to 10 photons with a detection efficiency exceeding 95\%~\cite{lita2008counting}.
The interferometer is a fiber-based device in which we can control the distance between two evanescently-coupled fibers using a micrometer to vary $\phi$, much like changing the path length difference between two arms of an interferometer.
Further details on the experimental setup can be found in the Methods.

\begin{figure*}
    \centering
    \includegraphics[width=0.7\textwidth]{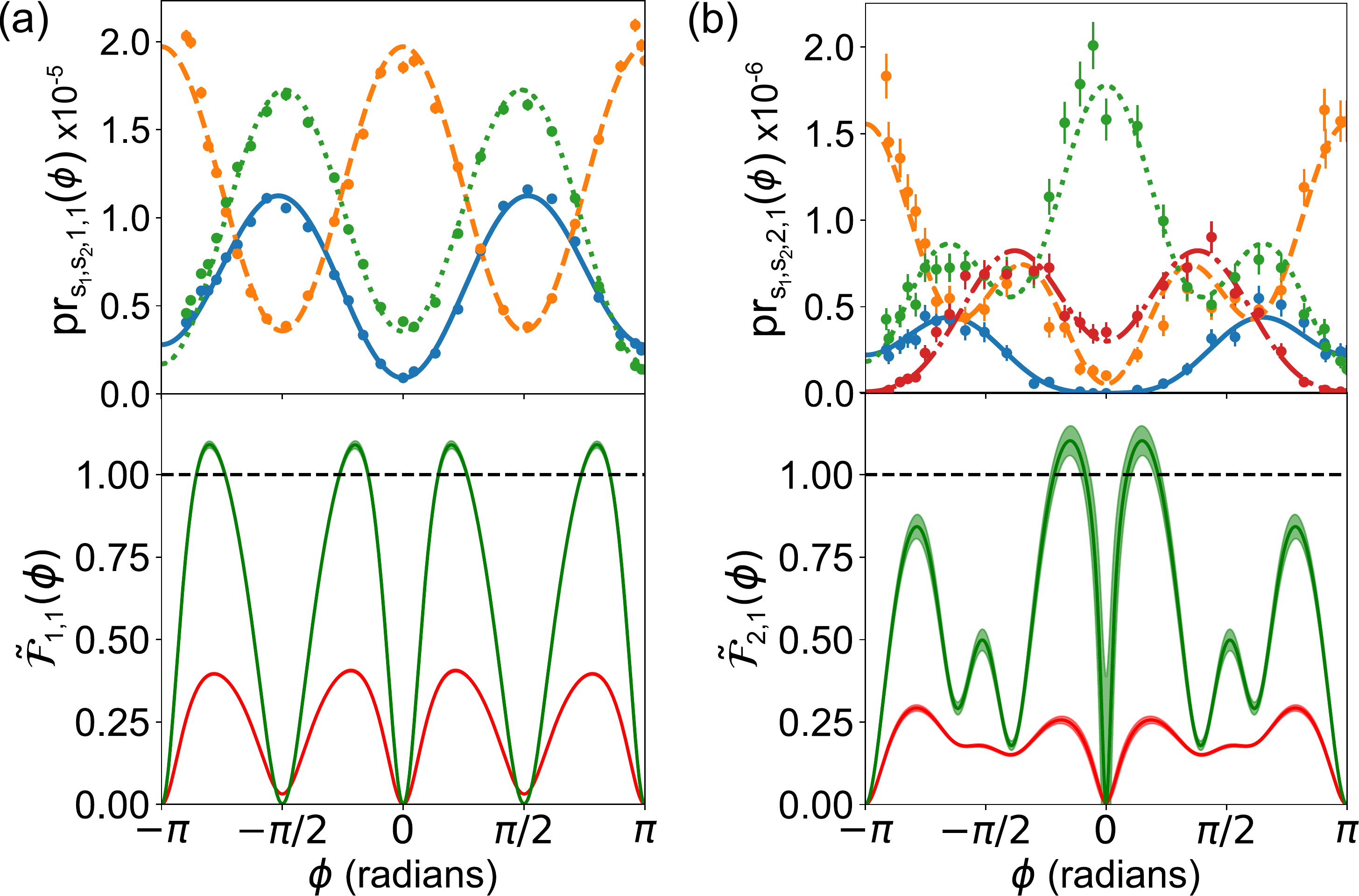}
	\caption{
    \textbf{The weak gain regime}.
    (a) Rates measured with the probe $\ket{1,1}$: $(s_1,s_2,h_1,h_2) = (2,0,1,1)$ [blue], $(1,1,1,1)$ [orange], $(0,2,1,1)$ [green].
    (b) Rates measured with the probe $\ket{2,1}$: $(3,0,2,1)$ [blue], $(2,1,2,1)$ [orange], $(1,2,2,1)$ [green], $(0,3,2,1)$ [red].
    Error bars are one standard deviation assuming Poissonian counting statistics.
    Lines are a model fitted to $\mathrm{pr}_{s_1,s_2,h_1,h_2}(\phi)$.
    Bottom panels show the normalized Fisher information $\tilde{\mathcal{F}}_{h_1,h_2}(\phi)$ calculated using two methods: (i) post-selecting on events where $s_1 + s_2 = h_1+h_2$ [green] and (ii) using all events [red].
    Line thicknesses show $1\sigma$ confidence intervals obtained by fitting 50 simulated data sets that are calculated with a Monte Carlo method.
    The dashed black line indicates the shot-noise limit.
	}
	\label{fig:lowSq}
\end{figure*}

We measure interference fringes given by $\mathrm{pr}_{s_1,s_2,h_1,h_2}(\phi)$, the joint photon-number probability per pump pulse to obtain the herald outcome $(h_1, h_2)$ and measure $(s_1,s_2)$ at the output of the interferometer when the phase difference is $\phi$.
We will refer to this as the $(s_1,s_2,h_1,h_2)$ rate.
To quantify the phase sensitivity of the rates measured with a particular herald outcome $(h_1, h_2)$, we calculate the Fisher information:
\begin{equation}
    \mathcal{F}_{h_1,h_2}(\phi) = \sum_{s_1,s_2} \frac{ \left[ \partial_\phi \tilde{\mathrm{pr}}_{s_1,s_2,h_1,h_2}(\phi)\right]^2}{\tilde{\mathrm{pr}}_{s_1,s_2,h_1,h_2}(\phi)},
    \label{eqn:cfi}
\end{equation}
where $\partial_\phi$ denotes the partial derivative with respect to $\phi$, and $\tilde{\mathrm{pr}}_{s_1,s_2,h_1,h_2}(\phi)$ is a model fitted to the measured rates (see Supplementary Method 2).
Note that $\mathcal{F}_{h_1,h_2}(\phi)$ quantifies the amount of information about $\phi$ in our measurement results, i.e. for a specific measurement strategy, and so $\mathcal{F}_{h_1,h_2}(\phi) \leq \mathcal{Q}$.
We compare the performance of our photon counting strategy to the optimal measurement strategy in the Supplementary Discussion 1.

Our primary figure of merit is the Fisher information per detected signal photon conditioned on measuring $(h_1,h_2)$ at the heralding detectors,
$$\tilde{\mathcal{F}}_{h_1,h_2}(\phi) = \mathcal{F}_{h_1,h_2}(\phi) / \braket{\tilde{n}}, $$
where
\begin{equation}
   \braket{\tilde{n}} = \sum_{s_1,s_2} (s_1+s_2)\tilde{\mathrm{pr}}_{s_1,s_2,h_1,h_2}(\phi)
   \label{eqn:avg_n}
\end{equation}
is the total number of detected signal photons. 
Injecting a coherent state into our interferometer would in principle yield the Fisher information $\mathcal{F}=\braket{\tilde{n}}$ when the detected mean photon number is $\braket{\tilde{n}}$~\cite{datta2011quantum}.
Thus, our figure of merit can be easily compared to the shot-noise limit which corresponds to $\tilde{\mathcal{F}}_{h_1,h_2}(\phi)=1$.

We measured the total efficiency of both the heralding and signal modes to be between $47-55 \%$ (see Supplementary Method 3).
This includes $\sim 90\%$ waveguide transmission, $\sim 70\%$ mode coupling efficiency into fibers, 90\% interferometer transmission, and $\gtrsim 95\%$ detector efficiency.
Due to the latter two losses, the detected $\braket{\tilde{n}}$ is 10-15\% smaller than the mean photon number inside the interferometer.
As such, the Fisher information per photon inside the interferometer (which is the relevant resource when e.g. probing a delicate sample) is about 10-15\% smaller than $\tilde{\mathcal{F}}_{h_1,h_2}(\phi)$.

\section*{Results}

We begin with low pump power to test our setup in the weak gain regime ($\lambda \sim 0.25$, $10~\mu$W per source).
In Fig.~\ref{fig:lowSq}, we show results for two different probes, (a) $\ket{1,1}$, the well-studied $N=2$ N00N or Holland-Burnett state, and (b) $\ket{2,1}$, a probe studied here for the first time.
We calculate $\tilde{\mathcal{F}}_{h_1,h_2}(\phi)$ using two methods.
In the first, we discard events in which we know photons were lost by only including rates where $s_1+s_2 = h_1+h_2$ in the sums of Eqs.~\eqref{eqn:cfi} and~\eqref{eqn:avg_n}.
These rates are shown in the top panels of Fig.~\ref{fig:lowSq}.
Using this first method, $\tilde{\mathcal{F}}_{h_1,h_2}(\phi)$ [green curves] surpasses the shot-noise limit by $0.09 \pm 0.01$ for $\ket{1,1}$ and $0.10 \pm 0.04$ for $\ket{2,1}$ at its highest point.
In the second method, we include all measured events.
Note that this may include events where $s_1 + s_2 < h_1+h_2$ due to loss in the signal modes, but also $s_1 + s_2 > h_1+h_2$ due to loss in the herald modes.
Conditioned on obtaining the herald outcome $(h_1,h_2)$, the probability of the latter occurring can be minimized by reducing the pump power and hence $\lambda$.
This increases the purity of the probe at the cost of reducing its heralding rate.
Without post-selection, $\tilde{\mathcal{F}}(\phi)$ [red curves] drops below the shot-noise limit mainly due to losses.

\begin{figure*}
    \centering
    \includegraphics[width=1\textwidth]{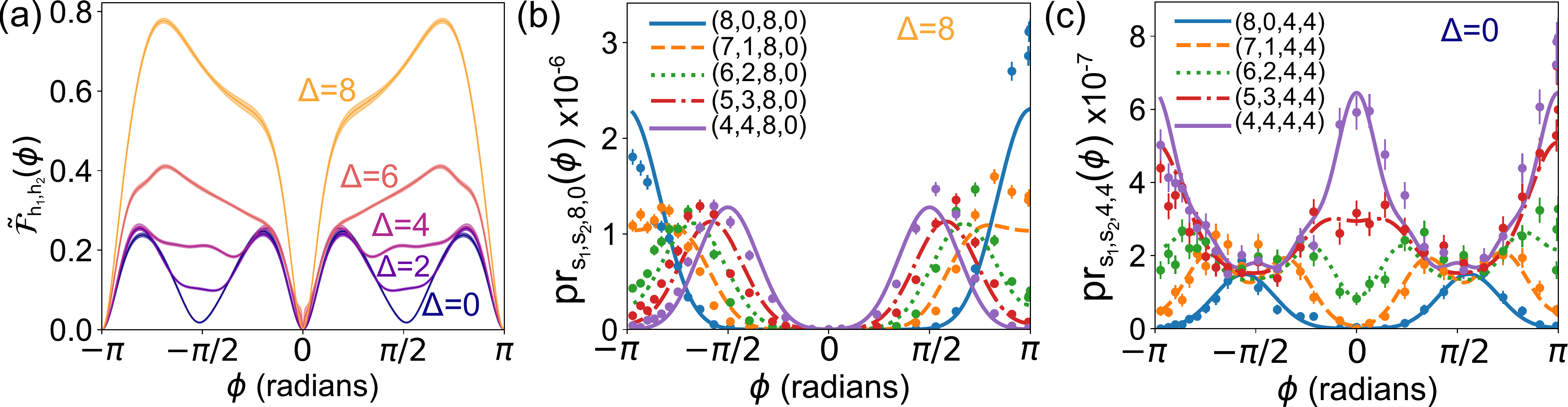}
	\caption{
    \textbf{The high gain regime}.
    (a) $\tilde{\mathcal{F}}_{h_1,h_2}(\phi)$ of 8-photon probes ($N=8$) parameterized by $\Delta = |h_1-h_2|$.
    Curves are calculated using the data and Eqs.~\eqref{eqn:cfi} and~\eqref{eqn:avg_n} without post-selection.
    Probes with a larger $\Delta$ have a larger $\tilde{\mathcal{F}}_{h_1,h_2}(\phi)$ and hence greater phase sensitivity due to their increased robustness to loss.
    Line thicknesses show $1\sigma$ confidence intervals obtained by fitting 50 simulated data sets that are calculated with a Monte Carlo method.
    (b) and (c) show a subset of rates for the probe with $\Delta = 8$ and $\Delta=0$, respectively.
    Error bars are one standard deviation assuming Poissonian counting statistics.
    The lines are a model fitted to $\mathrm{pr}_{s_1,s_2,h_1,h_2}(\phi)$.
    }
	\label{fig:highSqCFI}

\end{figure*}

In addition to loss, the spectral purity and distinguishability of our photons are also sources of imperfection that reduce the contrast of the fringes and hence diminish $\tilde{\mathcal{F}}_{h_1,h_2}(\phi)$~\cite{birchall2016beating}.
Consider the probe $\ket{1,1}$, for example.
For $\phi = \pm \pi/2$, the whole interferometer acts as a balanced beam splitter, in which case Hong-Ou-Mandel interference should lead to a complete suppression in coincidences at its output.
However, as can be seen in the orange $(1,1,1,1)$ fit in Fig.~\ref{fig:lowSq}(a), the visibility of this interference effect is $\sim 75\%$.
This visibility exceeds $\sqrt{0.5}$, which is the minimum required for demonstrating post-selected quantum-enhanced sensitivity with the probe $\ket{1,1}$~\cite{nagata2007beating,resch2008timereversal,thomas2011realworld}.
In addition to spectral mismatch between the signal modes, the visibility is degraded by uncorrelated background photons ($\sim 5\%$ of detected photons) and the slight multi-mode nature of our sources, both of which reduce the purity of our heralded photons.
We discuss source imperfections in more detail in the Supplementary Discussion 2.
The finite detector energy resolution also plays a small role as the detectors have a $\sim 1 \%$ chance to mislabel an event by $\pm 1$ photon~\cite{humphreys2015tomography}.

Next, we increase the pump power to reach a high-gain regime ($\lambda \sim 0.75$, $135~\mu$W per source) in which we can herald large photon numbers.
We detect 16-photon events at a rate of roughly $7$ per second, which is much higher than the state-of-the-art achievable with bulk crystal PDC sources~\cite{wang2016experimental} or quantum dots~\cite{wang2019boson}.
In Fig.~\ref{fig:highSqCFI}(a), we plot $\tilde{\mathcal{F}}_{h_1,h_2}(\phi)$ calculated without post-selection for all probes with $N=8$.
As expected given the amount of loss in our experiment, probes with larger $\Delta$ are more phase sensitive due to their increased robustness to loss [Fig.~\ref{fig:Scheme}(b)].
In particular, the sensitivity of the $\Delta=N$ probe should be shot-noise limited regardless of losses~\cite{pezze2007phase}.
However in practice, the heralded detection of $0$ photons could occur due to photon loss in the corresponding herald mode, resulting in the contamination of the signal with states for which $\Delta\ne N$.
This degrades the performance of the $\Delta=8$ probe [orange curve].
In the Supplementary Discussion 3, we show that shot-noise limited performance with the $\Delta=N$ probe is recovered by blocking one of the sources.

\section*{Discussion}
\begin{figure}
    \centering
    \includegraphics[width=1\columnwidth]{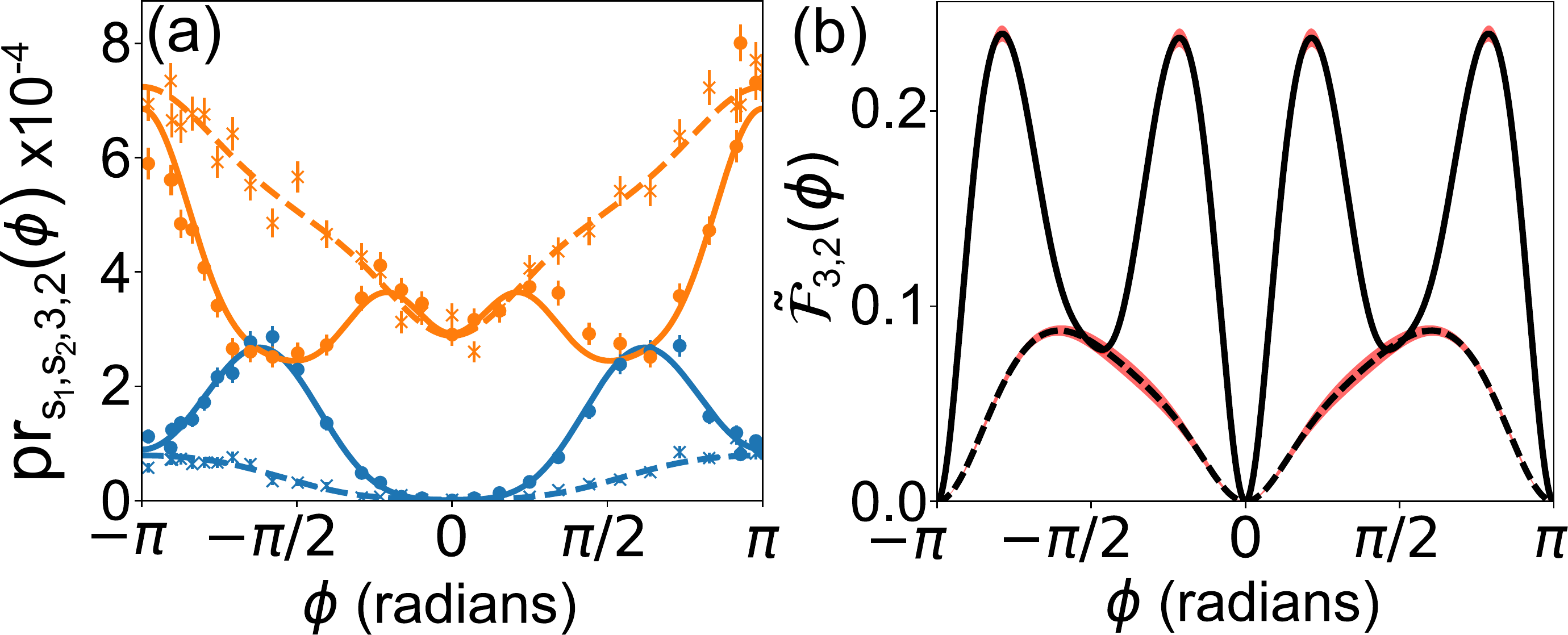}
	\caption{
    \textbf{Testing multiphoton interference}.
    Benefits of multiphoton interference using the probe $\ket{3,2}$. 
    (a) Two sets of rates [blue: $(5,0,3,2)$, orange: $(3,2,3,2)$] measured when the photons are injected inside the interferometer at the same time (data: circles, theory: bold lines) or at different times (data: crosses, theory: dashed line).
    In the latter case, the photons are well modelled by classical distinguishable particles.
    Error bars are one standard deviation assuming Poissonian counting statistics.
    (b) $\tilde{\mathcal{F}}_{3,2}(\phi)$ shows a significant improvement in sensitivity in the former case (bold line) compared to the latter case (dashed line), demonstrating that multiphoton interference improved the sensitivity of our probe.
    Red shaded regions shows $1\sigma$ confidence intervals obtained by fitting 50 simulated data sets that are calculated with a Monte Carlo method.
	}
	\label{fig:effect_of_distinguishability}
\end{figure}

The fringes produced by our probes exhibit a number of different features compared to those measured with N00N or Holland-Burnett states.
For example, with these two states, the expected signature of $N$-photon interference are fringe oscillations that vary as $\cos(N\phi)$.
While our measured fringes do not exhibit such oscillations in the high gain regime, they do exhibit sharper features than classical fringes.
We show this explicitly by comparing our rates to those measured with distinguishable photons.
This is achieved by temporally delaying photons coming from the top source with respect to photons coming from the bottom source by more than their coherence time.
As an example, we consider the probe $\ket{3,2}$ in Fig.~\ref{fig:effect_of_distinguishability}.
When the photons are injected inside the interferometer at the same time, the fringe contrast is significantly higher than when they are temporally delayed [Fig.~\ref{fig:effect_of_distinguishability}(a)].
Likewise, when we calculate $\tilde{\mathcal{F}}_{3,2}(\phi)$ without post-selection, we find an improvement in the probe's sensitivity in the former case [Fig.~\ref{fig:effect_of_distinguishability}(b)].
This demonstrates that the probe sensitivity derives from multiphoton interference even at high photon numbers.

With any finite amount of loss, $\tilde{\mathcal{F}}_{h_1,h_2}(\phi)$ vanishes when all fringes share a common turning point such as at $\phi = 0$.
In the case of Holland-Burnett ($\Delta = 0$) and N00N states, there are also common turning points at $\phi = \pm \pi/2$ which causes the reduction in $\tilde{\mathcal{F}}_{h_1,h_2}(\phi)$ around these phase values [Fig.~\ref{fig:highSqCFI}(c)]. 
In contrast, the probes with $\Delta=4,6,8$ do not have a dip in $\tilde{\mathcal{F}}_{h_1,h_2}(\pm \pi/2)$.
The origin of this effect for $\Delta = 8$ can be seen directly in the rates shown in Fig.~\ref{fig:highSqCFI}(b).
The region of the fringe with high sensitivity to $\phi$ (i.e. large gradient) is different for different values of $|s_1-s_2|$.
This feature of $\tilde{\mathcal{F}}_{h_1,h_2}(\phi)$ allows estimating $\phi$ without prior knowledge of the range in which it lies, as is required for N00N or Holland-Burnett states, and thus provides a means for global phase estimation without using an adaptive protocol~\cite{xiang2011entanglement,daryanoosh2018adaptive}.

Finally, we briefly compare our results to other works reporting Fisher information per detected photon.
The highest achieved here is $\sim 1.1$ using the herald outcome $(2,1)$, i.e. a $N=3$ probe.
Ref.~\cite{matthews2016towards} and Ref.~\cite{slussarenko2017unconditional} respectively report $\sim 1.25$ and $\sim 1.2$ using a $N=2$ probe.
The latter work also achieves a Fisher information per photon inside the interferometer (i.e. accounting for undetected photons) of $~\sim 1.15$ which thus far is the only experiment demonstrating an unconditional improvement to the shot-noise limit.
In the Supplementary Discussion 4, we estimate that an efficiency of $80\%$ (in all four modes) and quantum interference visibility of $85\%$ would be sufficient to demonstrate an improvement to the shot-noise limit with $N=8$ photons without post-selection.
Although we do not attain these parameters in our experiment, our results do demonstrate the robustness of our probes to losses despite their large size.
For example, the Fisher information per photon calculated without post-selection for the $N=8$ probe with $\Delta=6$ [Fig.~\ref{fig:highSqCFI}(a)] is slightly higher than that of the $N=2$ N00N state [Fig.~\ref{fig:lowSq}(a)].
This contradicts the usual expectation that large entangled probes will necessarily be more fragile to noise and loss.

In summary, we proposed and experimentally demonstrated a scheme for quantum-enhanced interferometry that exploits bright two-mode squeezed vacuum sources and photon-number-resolving detectors.
We measured interference fringes involving up to 16 photons which is significantly higher than the previous state-of-the-art~\cite{gao2010experimental,wang2016experimental}.
Crucially, our scheme prepares probes that are nearly optimally robust to losses and hence addresses one of the principal challenges when scaling-up to large entangled photonic states.
With further improvements in the quality (e.g. coupling efficiency into optical fiber and purity) of bright two-mode squeezed vacuum sources compatible with transition edge sensors~\cite{harder2016mesoscopic,vaidya2019broadband}, we believe our loss-tolerant scheme provides a promising route towards achieving quantum-enhanced resolution using large entangled photonic states.

\section*{Methods}

\subsection*{Sources}
We pick 150-fs pulses from a mode-locked Ti:Sapphire laser (Coherent Mira-HP) at a rate of 100 kHz using a Pockels-cell-based pulse picker having a 50 dB extinction ratio.
This repetition rate is chosen to accommodate the recovery time of the transition edge sensor detectors.
The pump pulses are filtered to $783$ $\pm~2$ nm  [full-width at half maximum] using a pair of angle-tuned bandpass filters. 
We split the pulses into two paths that are matched in length using a translation stage.
In each path, we pump a 8 mm long ppKTP waveguide that is phase-matched for type-II parametric down-conversion.
At the exit of the waveguide, the pump light is rejected with a longpass filter, and the orthogonally-polarized down-converted modes are separated using a polarizing beam splitter. 
Each down-converted mode is filtered with a bandpass filter whose bandwidth is chosen to transmit the main feature of the down-converted spectrum but reject its side-lobes.
The herald modes ($1566$ $\pm~7$ nm) are coupled into single-mode fibers and sent directly to the detectors.
The signal modes ($1567$ $\pm~7$ nm) are coupled into polarization-maintaining single-mode fibers and sent into the interferometer.
Details on the coupling efficiency and the spectral indistinguishability of the signal modes are provided in the Supplementary Discussion 2.
 
\subsection*{Interferometer}
The interferometer is a fiber-based variable beam splitter (Newport F-CPL-1550-P-FP).
The splitting ratio is adjusted by controlling the distance between two evanescently-coupled fibers using a micrometer, which is analogous to changing the path length difference between two arms of an interferometer.
In fact, any variable beam splitter that coherently splits light into two modes can be described by the same transformation as a Mach-Zender-type interferometer~\cite{florez2018variable}. 

During data acquisition, we scan the distance $x$ between the two evanescently-coupled fibers.
To display our data as a function of the interferometer phase, we first calculate the transmission coefficient $T(x)$ of the variable beam splitter using the measured $(1,0,1,0)$ and $(0,1,1,0)$ rates:
\begin{equation}
    T(x) = \frac{\mathrm{pr}_{1,0,1,0}(x)}{\mathrm{pr}_{1,0,1,0}(x) + \mathrm{pr}_{0,1,1,0}(x)}.
\end{equation}
At low powers, we find that the quantity $T(x)$ typically varies within $[0.02, 0.98]$. %, indicating that the interferometer visibility is at least $96\%$.
To obtain the corresponding phase, we correct for the imperfect visibility:
\begin{equation}
    T_{\mathrm{corr}}(x) = \frac{T(x) - \min{[T(x)]}}{\max{[T(x)]}-\min{[T(x)]}} 
\end{equation}
such that $T_{\mathrm{corr}}(x)$ varies between [0,1].
For a single photon injected into a Mach-Zender type interferometer with phase difference $\phi$ between its two arms, one expects $T_{\mathrm{corr}}(x) = [1-\cos{(\phi)}]/2$.
Solving for $\phi$, we find:
\begin{equation}
\phi(x) = \arccos{\left(2T_{\mathrm{corr}}(x)-1\right)}.
\end{equation}

\subsection*{Detectors}
Our detectors are superconducting transition edge sensor detectors that operate at a temperature of 85 mK inside a dilution refrigerator.
Details on their physical operation can be found in Ref.~\cite{lita2008counting}.
An electrical trigger signal from the pump laser begins a 6 $\mu s$ time window of data acquisition during which the detector outputs are amplified and recorded with an analogue-to-digital converter.
We use a matched-filter technique in real-time to convert each detector's output trace into a scalar value~\cite{figueroa2000optimal}.
The scalar value is then converted into a photon number using bins that are set during an initial calibration run prior to data acquisition.

\section*{Data availability}
The data sets generated and/or analyzed during this study are available from the corresponding author on reasonable request.
Correspondence and requests for materials should be addressed to G.S.T.$^{*}$

\section*{Acknowledgements}
We thank B. Vlastakis for his assistance with the operation of the dilution refrigerator.
This work was supported by the following: the Natural Sciences and Engineering Research Council of Canada (NSERC); the Networked Quantum Information Technologies Hub (NQIT) as part of the UK National Quantum Technologies Programme GrantEP/N509711/1) and within ``First
Team" project No. POIR.04.04.00-00-220E/16-00 (originally: FIRST TEAM/2016-2/17) of the Foundation for Polish Science co-financed by the European Union under the European Regional Development Fund.

\section*{Competing interests}
The authors declare no competing interests.

\section*{Author contributions}
Both G.S.T. and M.E.M. contributed equally.
G.S.T. performed the experiment with assistance from B.A.B, C.G.W, A.E, D.S.P.;
M.E.M. and G.S.T. performed numerical calculations with assistance from A.B.;
A.E.L., T.G., and S.W.N. developed the detectors;
R.B.P., M.S., A.I.L., and I.A.W. initiated and/or supervised the project;
G.S.T. and M.E.M. wrote the manuscript with input from all authors.

\bibliographystyle{apsrev4-1}
\bibliography{refs}

%%% SM %%%

% UNCOMMENT FOR ARXIV

\newpage
\onecolumngrid

\section*{Supplementary information}

\renewcommand{\theequation}{S\arabic{equation}}
\renewcommand{\thefigure}{S\arabic{figure}}

\subsection*{Supplementary Note 1: Quantum Fisher information of generalized Holland-Burnett states}

Here we derive the quantum Fisher information of our generalized Holland-Burnett states.
We first consider the ideal lossless case.
In general, the quantum Fisher information of a pure state $\ket{\Psi(\phi)}$ that depends on some parameter $\phi$ is given by~\cite{braunstein1994statistical}:
\begin{equation}
    \mathcal{Q} = 4\left( \braket{\partial_\phi\Psi(\phi)|\partial_\phi\Psi(\phi)} - \left| \braket{\partial_\phi\Psi(\phi)|\Psi(\phi)} \right|^2 \right)
\end{equation}
where $\ket{\partial_\phi\Psi(\phi)} \equiv \partial_\phi \ket{\Psi(\phi)}$.
In our case, $\ket{\Psi}$ is the two-mode state inside the interferometer (before the phase shift) and $\ket{\Psi(\phi)}=e^{i\hat{c}^\dagger\hat{c}\phi}\ket{\Psi}$ is the state after the phase shift $\phi$ is applied in the upper interferometer mode $c$.
After some simple algebra, one finds that $\mathcal{Q}$ is independent of $\phi$ and is determined by:
\begin{equation}
\mathcal{Q} = 4 \left[\braket{\Psi | (\hat{c}^\dagger\hat{c})^2 | \Psi} - \braket{\Psi | \hat{c}^\dagger\hat{c}| \Psi}^2\right].
\label{eqn:QFI}
\end{equation}
%producing the  phase rotation in mode $a$, i.e. the upper mode inside the interferometer. 
%Here $a \, (a^{\dagger})$ is the annihilation (creation) operator in mode $a$.
%measuring the acquired optical phase in one of the modes. This phase is proportional to the photon number in that mode. In general, $H = a^{\dagger}a$ where $a \, (a^{\dagger})$ is the annihilation (creation) operator in mode $a$.

We wish to calculate $\mathcal{Q}$ for the particular probe $\ket{\Psi}=\hat{U}_{\mathrm{BS}}\ket{h_1,h_2}$ where $\hat{U}_{\mathrm{BS}}$ is the balanced beam splitter unitary transformation.
The second term in Eq.~\eqref{eqn:QFI} is given by:
\begin{align}
\braket{\Psi| \hat{c}^\dagger\hat{c} |\Psi} &= \braket{h_1,h_2| \hat{U}_{\mathrm{BS}}^{\dagger}c^{\dagger}c \, \hat{U}_{\mathrm{BS}} |h_1,h_2}\\
&= \braket{h_1,h_2| \hat{U}_{\mathrm{BS}}^{\dagger}c^{\dagger}\hat{U}_{\mathrm{BS}}\,\hat{U}^{\dagger}_{\mathrm{BS}}c \, \hat{U}_{\mathrm{BS}} |h_1,h_2}\label{unitary}\\
&= \braket{h_1,h_2| \left(\frac{a^{\dagger} + b^{\dagger}}{\sqrt{2}}\right) \left(\frac{a+b}{\sqrt{2}}\right) |h_1,h_2}\label{BS_transformation}\\
&= \frac{1}{2}\braket{h_1,h_2| a^{\dagger}a + b^{\dagger}b |h_1,h_2}\\
&= \tfrac{1}{2} (h_1+h_2),
\end{align}
where in line~\eqref{unitary} we used the fact that $\hat{U}_{\mathrm{BS}}$ is unitary and in line~\eqref{BS_transformation} we transformed mode $c$ to the input modes $a$ and $b$.
The first term in equation~\eqref{eqn:QFI} is calculated in a similar manner:
\begin{align}
\braket{\Psi| (\hat{c}^\dagger\hat{c})^2 |\Psi} &= \braket{h_1,h_2| \hat{U}_{\mathrm{BS}}^{\dagger}c^{\dagger}c c^{\dagger}c \, \hat{U}_{\mathrm{BS}} |h_1,h_2}\\
&= \braket{h_1,h_2|\left( \hat{U}_{\mathrm{BS}}^{\dagger}c^{\dagger}\hat{U}_{\mathrm{BS}}\,\hat{U}^{\dagger}_{\mathrm{BS}}c \, \hat{U}_{\mathrm{BS}}\right)^2 |h_1,h_2}\\
&= \braket{h_1,h_2| \left(\frac{a^{\dagger} + b^{\dagger}}{\sqrt{2}}\right)^2 \left(\frac{a+b}{\sqrt{2}}\right)^2 |h_1,h_2}\\
&= \frac{1}{4}\braket{h_1,h_2| (a^{\dagger} + b^{\dagger})^2 (a+b)^2 |h_1,h_2}\\
&= \frac{1}{4}\braket{h_1,h_2| a^{\dagger}aa^{\dagger}a + b^{\dagger}bb^{\dagger}b + 4a^{\dagger}ab^{\dagger}b + a^{\dagger}a + b^{\dagger}b |h_1,h_2}\\
&= \tfrac{1}{4}(h_1^2 + h_2^2 + 4h_1h_2 + h_1 + h_2).
\end{align}
Therefore $\mathcal{Q}$ is given by
\begin{equation}
\mathcal{Q} = \tfrac{1}{4}(h_1^2 + h_2^2 + 4h_1h_2 + h_1 + h_2) - \tfrac{1}{4} (h_1+h_2)^2 = 2h_1h_2 + h_1 + h_2.
\label{eqn:QFI_gHB}
\end{equation}

Eq.~\eqref{eqn:QFI_gHB} only applies when there are no losses in the system. 
In the presence of losses, the probe $\ket{\Psi}$ is transformed to a mixed state $\hat{\rho}$.
Then, $\mathcal{Q}$ is calculated using
\begin{equation}
    \mathcal{Q} = \mathrm{Tr}\Big\{\hat{\rho}(\phi)\Lambda^2[\hat{\rho}(\phi)]\Big\}
    \label{eqn:QFI_general}
\end{equation}
where $\hat{\rho}(\phi) = e^{-i\hat{c}^\dagger\hat{c}\phi}\hat{\rho}e^{i\hat{c}^\dagger\hat{c}\phi}$ is the probe state after the phase shift $\phi$ and $\hat{\Lambda}[\hat{\rho}(\phi)]$ is a Hermitian operator called the ``symmetric logarithmic derivative" defined implicitly via
\begin{equation}
    \partial_\phi \hat{\rho}(\phi) = \frac{1}{2}\Big\{\hat{\Lambda}[\hat{\rho}(\phi)]\hat{\rho}(\phi) + \hat{\rho}(\phi)\hat{\Lambda}[\hat{\rho}(\phi)]\Big\}.
    \label{eqn:implicit_SLD}
\end{equation}
We notice that by combining Eq.~\eqref{eqn:implicit_SLD} with Eq.~\eqref{eqn:QFI_general} we obtain an alternative equation for the QFI
\begin{equation}
    \mathcal{Q} = \mathrm{Tr}\Big\{\partial_{\phi}\hat{\rho}(\phi)\Lambda[\hat{\rho}(\phi)]\Big\}.
\end{equation}
By writing $\hat{\rho}$ in its eigenbasis, $\hat{\rho}=\sum_i p_i\ket{e_i}\bra{e_i}$ and writing out the derivative $\partial_{\phi}\hat{\rho}(\phi) = ie^{-i\hat{c}^\dagger\hat{c}\phi}[\hat{\rho}, \hat{c}^\dagger\hat{c}]e^{i\hat{c}^\dagger\hat{c}\phi}$, it can be shown that $\mathcal{Q}$ is given by~\cite{RDD2015review}
\begin{equation}
    \mathcal{Q} = \sum_{i,j} \frac{2\left|\braket{e_i|\hat{c}^\dagger\hat{c}|e_j}\right|^2(p_i-p_j)^2}{p_i+p_j}
    \label{eqn:qfi_mixed_state}
\end{equation}
which is independent of $\phi$.
The sum is taken over all terms with a non-vanishing denominator.
% and using:
% \begin{equation}
% \mathcal{Q} = \mathrm{Tr}[\hat{\rho} \hat{\Lambda}^2],
% \label{eqn:qfi_mixed}
% \end{equation}
% where $\hat{\Lambda}$ is an Hermitian operator called the ``symmetric logarithmic derivative'' (SLD) and is defined implicitly via
% \begin{equation}
% \partial_\phi\hat{\rho}(\phi) = \frac{1}{2} [\hat{\Lambda} \hat{\rho}(\phi) + \hat{\rho}(\phi) \hat{\Lambda}].
% \label{eqn:sld}
% \end{equation}
% In the eigenbasis $\hat{\rho}(\phi) = \sum_i p_i \ket{e_i}\bra{e_i}$, $\hat{\Lambda}$ can be calculated directly
% \begin{equation}
% \hat{\Lambda} = \sum_{i,j} \frac{2 \bra{e_i}\partial_\phi\hat{\rho}(\phi){\ket{e_j}}}{p_i + p_j} \ket{e_i}\bra{e_j},
% \end{equation}
% where the sum is taken over all terms with a non-vanishing denominator.
% Note that 

\subsection*{Supplementary Method 1: Optimal states}
In the main text we compare the performance of our probes to the ``optimal states" which provide the largest possible quantum Fisher information given some amount of loss~\cite{dorner2009optimal,RDD2009optimal}.
A general $N$-photon pure state inside the interferometer can be written in the Fock basis as
\begin{equation}
\ket{\Psi} = \sum_{n=0}^N \alpha_n \ket{n, N-n}.
\label{eq:general_state}
\end{equation}
In the absence of loss, the optimal state is found by optimizing the coefficients $\{ \alpha_n \}$ to maximize the quantum Fisher information $\mathcal{Q} = 4 \left[\braket{\Psi | (\hat{c}^\dagger\hat{c})^2 | \Psi} - \braket{\Psi | \hat{c}^\dagger\hat{c}| \Psi}^2\right]$.

In the presence of loss, $\ket{\Psi}$ turns into a mixture $\hat{\rho}$ which can be written in the following form
\begin{equation}
\hat{\rho} = \sum_j p_j \ket{\Psi_j}\bra{\Psi_j},
\end{equation}
where $\ket{\Psi_j}$ do not have to be orthogonal. 
Due to the convexity of quantum Fisher information, $\mathcal{Q}'$ of $\hat{\rho}$ is upper bounded by
\begin{equation}
\mathcal{Q}' \leq \mathcal{Q} = 4 \sum_j p_j \left(\braket{\Psi_j| (\hat{c}^\dagger\hat{c})^2 | \Psi_j} - \braket{\Psi_j |\hat{c}^\dagger\hat{c} | \Psi_j}^2\right).
\label{eq:QFI_bound}
\end{equation}
The bound is attained if the kets $\ket{\Psi_j}$ are orthogonal, which is the case for e.g. N00N states or if photon losses are present in only one interferometer mode.

Applying Eq.~\eqref{eq:QFI_bound} to Eq.~\eqref{eq:general_state}, we obtain
\begin{equation}
\mathcal{Q} = 4 \bigg( \sum_{n=0}^N n^2 x_n - \sum_{l=0}^N \sum_{m=0}^{N-l} \frac{\big(\sum_{k=l}^{N-m} x_n n B_{lm}^n\big)^2}{\sum_{n=l}^{N-m} x_n B_{lm}^n}\bigg),
\label{eqn:optimal}
\end{equation}
where $x_n = |\alpha_n|^2$, $B_{lm}^n \equiv \binom{n}{l}\binom{N-n}{m}\eta_{s_1}^n(\eta_{s_1}^{-1} - 1)^l \eta_{s_2}^{N-n}(\eta_{s_2}^{-1} - 1)^m$ and $\eta_{s_1}, \eta_{s_2}$ denote the transmittances in the signal modes.

The optimal states are found by numerically maximizing $\mathcal{Q}$ over the probabilities $\{x_n\}$. 
Since $\mathcal{Q}$ is a concave function of $\{x_n\}$~\cite{RDD2009optimal}, any maximum is global. 
Although $\mathcal{Q}' < \mathcal{Q}$ [Eq.~\eqref{eq:QFI_bound}] when losses are present in both modes, the difference between the two quantities is small relative to the difference between the shot-noise limit and the Heisenberg limit~\cite{RDD2009optimal}.
Due to this approximation, the optimized $\mathcal{Q}$ is a slight over-estimate of true quantum Fisher information $\mathcal{Q}'$ of the optimal states.

Fig.~1(b) in the main text shows $\mathcal{Q}$ of our probes and the optimal state as a function of equal transmissivity in the signal modes $\eta_{s_1} = \eta_{s_2} = \eta_s$ which varied from 0 to 1 in steps of 0.01.
The optimal state was calculated in \textsc{Mathematica} by maximizing over coefficients $\{x_n\}$ in Eq.~\eqref{eqn:optimal}, assuming they all sum up to 1 and are real and positive.
We computed $\mathcal{Q}$ of our probes in Python using the following method.
We started with two copies of the state in Eq.~(1) in the main text, inserted a beam splitter in each signal mode, and traced over the reflected port to model signal transmissivities $\eta_s$. 
The two matrices were then combined on the first interferometer beam splitter forming a four mode density matrix, which was then reduced to two modes by projectively measuring the two herald modes. 
Eigenvalues and eigenvectors were found for the two-mode density matrix inside the interferometer which were then used to calculate $\mathcal{Q}$ via Eq.~\eqref{eqn:qfi_mixed_state}. 
To compare the ideal performance of our probes with the optimal state, we exclude the effect of imperfect heralding on the former by using $\eta_{h_1}=\eta_{h_2}=1$ in the calculation of $\mathcal{Q}$.

From the calculations described above we show that for $\eta_s \in (0, 0.5 \rangle$ the best approximation to the optimal state is given by the probe with $\Delta = 8$; for $\eta_s \in (0.5,0.58 \rangle$ by the probe with $\Delta = 6$; for $\eta_s \in (0.58,0.66 \rangle$ by the probe with $\Delta = 4$; for $\eta_s \in (0.66,0.69 \rangle$ by the probe with $\Delta = 2$; and for $\eta_s > 0.69$ by the probe with $\Delta = 0$, as shown by the colored line in Fig.~1(b).

\subsection*{Supplementary Method 2: Modelling the measured rates}

\begin{figure}
    \centering
    \includegraphics[width=0.75\columnwidth]{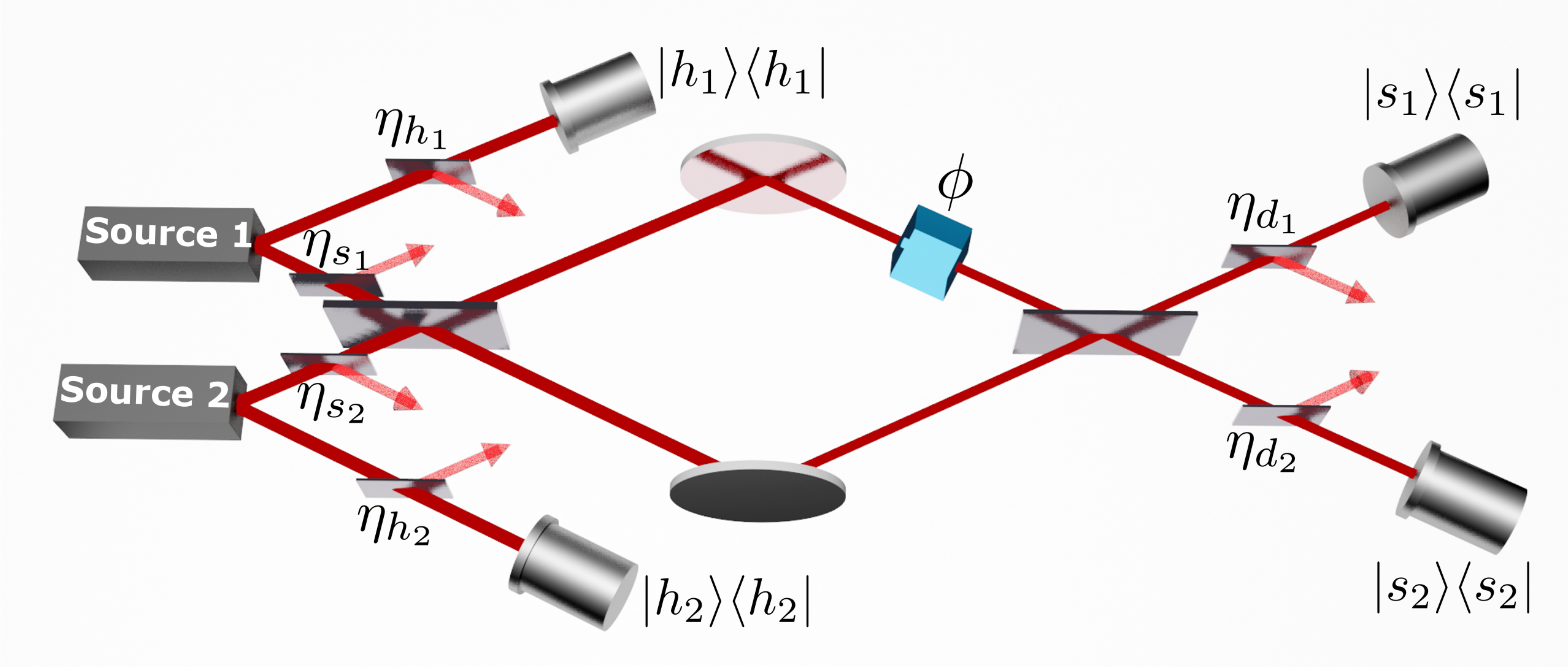}
	\caption{
    Losses are modelled by placing fictitious beam splitters in all four modes before the interferometer and just before the signal detectors.
    Coefficients show the transmission of the beam splitters.
    }
	\label{fig:schematic_loss}

\end{figure}

Here we describe the model $\tilde{\mathrm{pr}}(s_1,s_2,h_1,h_2,\phi)$ used to fit the experimentally measured rates.
We model optical loss by placing fictitious beam splitters (see Fig.~\ref{fig:schematic_loss}) and tracing over the reflected modes.
For now, we assume $\eta_{d_1}=\eta_{d_2}=1$.
We will treat the effect of these detection losses at the end.

The sources produce two-mode squeezed vacuum states:
\begin{equation}
    \ket{\chi_i} = \sqrt{1-\lambda_i^2}\sum_{n=0}^{\infty}\lambda_i^n\ket{n,n},
\end{equation}
where $i=1,2$ represent sources 1 and 2, respectively.
The joint photon-number distribution of this two-mode squeezed vacuum state after the losses is given by:
\begin{equation}
    \tilde{\mathrm{pr}}_i(x,y) = (1-\lambda_i^2)\sum_{n=\max{(x,y)}}^{\infty}{n \choose x}{n \choose y}\lambda^{2n}_i \eta_{h_i}^x \eta_{s_i}^y (1-\eta_{h_i})^{n-x} (1-\eta_{s_i})^{n-y}.
    \label{eqn:joint_photon_tmsv}
\end{equation}
The intuition for the expression above is as follows.
Imagine that there are two detectors after the fictitious beam splitters that give the detection outcome $(x,y)$.
The source must have produced at least $\max{(x,y)}$ photon pairs and perhaps some photons were lost (i.e. reflected at the beam splitters).
The probability to produce $n$ pairs of photons is $(1-\lambda_i)^2\lambda^{2n}_i$. 
Having produced $n$ pairs, the probability to reflect $n-x$ [$n-y$] photons and transmit $x$ [$y$] photons in the herald [signal] mode is ${n \choose x}\eta_{h_i}^x(1-\eta_{h_i})^{n-x}$ [${n \choose y}\eta_{s_i}^y(1-\eta_{s_i})^{n-y}$].
In principle, $n$ can range up to $\infty$, but in practice it suffices to truncate this sum at some value where $(1-\lambda_i)^2\lambda^{2n}_i$ becomes small.
In our numerics, we truncate the sum at $n=50$.

If we obtain the herald outcome $(h_1,h_2)$, then the (unnormalized) state that is injected into the interferometer is given by:
\begin{equation}
    \hat{\rho} = \sum_{m,n=0}^{\infty} \tilde{\mathrm{pr}}_1(h_1,m)\tilde{\mathrm{pr}}_2(h_2,n)\ket{m,n}\bra{m,n}.
    \label{eqn:input_state_interf}
\end{equation}
Losses occurring inside the interferometer can be absorbed into $\eta_s$ or $\eta_d$ if they are equal in both interferometer modes, which was approximately the case in our experiment.
Thus, the interferometer transformation can be described by a unitary operator $\hat{U}(\phi)$ which depends on the phase difference $\phi$ between both arms.
The probability that we wish to calculate is given by:
\begin{equation}
    \tilde{\mathrm{pr}}(s_1,s_2,h_1,h_2,\phi) = \braket{s_1,s_2|\hat{U}(\phi)\hat{\rho}\hat{U}^\dagger(\phi)|s_1,s_2}.
\end{equation}
Knowing that there are a total of $s_1+s_2$ photons before the interferometer, we can constrain $n=s_1+s_2-m$ and truncate the sum at $s_1+s_2$ in Eq.~\eqref{eqn:input_state_interf}.
Thus, we obtain:
\begin{equation}
    \tilde{\mathrm{pr}}(s_1,s_2,h_1,h_2,\phi) = \sum_{m=0}^{s_1+s_2}\tilde{\mathrm{pr}}_1(h_1,m)\tilde{\mathrm{pr}}_2(h_2,s_1+s_2-m) \left|\braket{s_1,s_2|\hat{U}(\phi)|m,s_1+s_2-m}\right|^2.
    \label{eqn:joint_prob_final}
\end{equation}
The matrix element $\left|\braket{s_1,s_2|\hat{U}(\phi)|m,s_1+s_2-m}\right|^2$ is derived in Ref.~\cite{leonhardt2010essential} and is given by:
% \begin{equation}
% \left|\braket{s_1,s_2|\hat{U}(\phi)|m,s_1+s_2-m}\right|^2 = \left|\phi_{s_1}^{(1/2)}\big(m - \tfrac{s_1 + s_2}{2}, s_1 + s_2 \big)\right|^2
%\end{equation}
\begin{equation}
\begin{split}
\left|\braket{s_1,s_2|\hat{U}(\phi)|m,s_1+s_2-m}\right|^2 &= \frac{m!(s_1+s_2-m)!}{s_1!s_2!}(\sin{[\phi/2]})^{2(s_1+m)}(\cos{[\phi/2]})^{2(s_2-m)} \\
&\qquad{}\times \left(\sum_{k=0}^{s_1} {s_1 \choose k} {s_2 \choose s_2+k-m} (-1)^{k} \tan{[\phi/2]}^{-2k} \right)^2.
\end{split}
\end{equation}
%\textcolor{red}{where $\phi_{s_1}^{(1/2)}$ is the Kravchuk function.}
Alternatively, the matrix element can also be evaluated using  Kravchuk polynomials~\cite{stobinska2019quantum}.

The model for temporally distinguishable photons follows the same approach as above.
While the derivation below focuses on temporal distinguishability, the same equations are valid to describe distinguishability in any other degree of freedom.
We adopt a heuristic approach (e.g. as in Ref~\cite{birchall2016beating}) in which the temporal mode of the photons produced in the top source is decomposed into a component completely indistinguishable ($\parallel$) to the temporal mode of the bottom source photons  as well as a component completely distinguishable ($\perp$). With this decomposition, Eq.~\eqref{eqn:input_state_interf} becomes:
\begin{equation}
    \hat{\rho}^{dist} =  \sum_{m,n=0}^{\infty} \sum_{l=0}^{m} {m \choose l} \mathcal{M}^{l} (1-\mathcal{M})^{m-l} \tilde{\mathrm{pr}}_1(h_1,m)\tilde{\mathrm{pr}}_2(h_2,n)\ket{l,n}_\parallel\bra{l,n}_\parallel \otimes \ket{m-l,0}_\perp\bra{m-l,0}_\perp.
    \label{eqn:input_state_interf_dist}
\end{equation}
where $\mathcal{M} \in [0,1]$ is a mode overlap parameter characterizing the distinguishability of the photons.
For $\mathcal{M} = 0$ ($\mathcal{M} = 1$), the photons from top and bottom sources are completely distinguishable (indistinguishable).
Since our detectors cannot resolve the time difference between $\perp$ and $\parallel$, they convolve the probabilities for the photons to have originated from either temporal mode.
This measurement is described by the following incoherent sum of projectors:
\begin{equation}
    \hat{\Pi} = \sum_{x=0}^{s_1}\sum_{y=0}^{s_2} \ket{s_1-x,s_2-y}_\parallel \bra{s_1-x,s_2-y}_\parallel \otimes \ket{x,y}_\perp \bra{x,y}_\perp.
    \label{eqn:povm_convolute}
\end{equation}
Many of the terms in the sum of Eq.~\eqref{eqn:povm_convolute} can be eliminated due to constraints on the photon numbers.
For example, a total of $m-l$ photons are produced in mode $\perp$ and so $x+y=m-l$.
Moreover, $s_1+s_2=m+n$.
After applying these constraints, the final joint probability is given by:
\begin{equation}
\begin{split}
\tilde{\mathrm{pr}}^{dist}(s_1,s_2,h_1,h_2,\phi) &= \mathrm{Tr}\left( \hat{\Pi} \hat{U}(\phi) \hat{\rho}^{dist} \hat{U}^\dagger(\phi) \right) \\
&= \sum_{m=0}^{s_1+s_2} \sum_{l=0}^m \sum_{x=\max(0, m-s_2)}^{\min(s_1,m-l)}  \binom{m}{l} \mathcal{M}^l(1-\mathcal{M})^{m-l} \tilde{\mathrm{pr}}_1(h_1,m)\tilde{\mathrm{pr}}_2(h_2,s_1+s_2-m) \\ 
&\qquad{}\times \left|\braket{s_1-x,l+s_2-m+x|\hat{U}(\phi)|l,s_1+s_2-m}\right|^2 \\
&\qquad{}\times \left|\braket{x,m-l-x|\hat{U}(\phi)|m-l,0}\right|^2.
% &= \sum_{m=0}^{s_1+s_2} \sum_{z=\max(0, m-s_2)}^{\min(s_1,m)}  \tilde{\mathrm{pr}}_1(h_1,m)\tilde{\mathrm{pr}}_2(h_2,s_1+s_2-m) \left|\phi_z^{(1/2)}\big(\tfrac{m}{2}, m\big)\right|^2 \\
% &\qquad{}\times \left|\phi_{s_1 - z}^{(1/2)}\big(-\tfrac{s_1 + s_2 - m}{2}, s_1 + s_2 - m\big)\right|^2
\end{split}
\label{eqn:joint_prob_final_dist}
\end{equation}

Finally, we can now consider the effect of the losses just before the detectors.
These losses can be modelled with a transformation analogous to Eq.~\eqref{eqn:joint_photon_tmsv}. 
Applying this transformation on Eq.~\eqref{eqn:joint_prob_final}, we obtain:
\begin{equation}
    \tilde{\mathrm{pr}}(s_1,s_2,h_1,h_2,\phi; \eta_{d_1}, \eta_{d_2}) = \sum_{j=s_1}^\infty\sum_{k=s_2}^\infty {j \choose s_1}{k \choose s_2} \eta_{d_1}^{s_1} \eta_{d_2}^{s_2} (1-\eta_{d_1})^{j-s_1} (1-\eta_{d_1})^{k-s_2} \tilde{\mathrm{pr}}(j,k,h_1,h_2,\phi)
    \label{eqn:final_joint_prob_wLoss}
\end{equation}
The same method is used for the distinguishable photons model, i.e. replace $ \tilde{\mathrm{pr}}(j,k,h_1,h_2,\phi)$ with $\tilde{\mathrm{pr}}^{dist}(j,k,h_1,h_2,\phi)$ in Eq.~\eqref{eqn:final_joint_prob_wLoss}.
In our numerics, we truncate the sums in Eq.~\eqref{eqn:final_joint_prob_wLoss} to only include the effect of losing a few photons, which is a good approximation given the high efficiency of our number-resolving detectors.

The equations above are evaluated numerically and fitted to the experimentally measured $\mathrm{pr}(s_1,s_2,h_1,h_2,\phi)$ by varying the fit parameters $\eta_{h_1},\eta_{h_2},\eta_{s_1},\eta_{s_2},\eta_{d_1},\eta_{d_2},\lambda_1,\lambda_2$.
Fitting is performed using the Python package \textsc{lmfit} with a least squares method.
Note that, for the sake of increasing the speed of the fitting, we used $\mathcal{M}=1$ for all data except for the dashed lines in Fig.~4 of the main text where we used $\mathcal{M}=0$.
Thus, the fit parameters generally did not correspond to the measured efficiencies and squeezing parameters (see below). 
Instead, the fitting procedure converged on larger $\lambda$ values and smaller $\eta$ values to emulate the effect of imperfect interference (i.e. reduced fringe visibility).
We tested the full model (i.e. including $\mathcal{M}$) by fitting a subset of rates measured in the high gain regime and found the fit parameters: $\eta_{h_1} = 0.50$, $\eta_{h_2} = 0.50$, $\eta_{s_1} = 0.61$, $\eta_{s_2} = 0.50$, $\eta_{d_1} = 0.9$, $\eta_{d_2} = 0.99$, $\lambda_1 = 0.68$, $\lambda_2 = 0.68$, and $\mathcal{M}=0.73$.
These efficiency values are within error to the measured values (see below), and $\mathcal{M}=0.73$ is roughly consistent with the measured $\sim 75 \%$ quantum interference visibility of the $(1,1,1,1)$ rate.

\subsection*{Supplementary Method 3: Estimating efficiencies}
We characterize the efficiency of our setup using a Klyshko-like method that is generalized to photon-number-resolving detection~\cite{worsley2009absolute}.
We set the variable beam splitter to maximize reflection and measure the joint photon-number distribution $\mathrm{pr}_i(x,y)$ pumping one source at a time.
We fit the measured $\mathrm{pr}_i(x,y)$ to $\tilde{\mathrm{pr}}_i(x,y)$ [see Eq.~\eqref{eqn:joint_photon_tmsv}] using three parameters: the PDC gain $\lambda_i$ and the total efficiency of the herald mode ($\eta_{h_i}$) and the signal mode ($\eta_{s_i}$).
Note that the latter will also include the detection efficiency $\eta_{d_i}$.
By repeating the procedure with five different pump powers, we find that $\eta_{s_1}\eta_{d_1}  = 56 \pm 3 \%$ and $\eta_{h_1} = 47 \pm 1 \%$ for the first source, and $\eta_{s_2}\eta_{d_2} = 52 \pm 4 \%$ and $\eta_{h_2} = 51 \pm 1 \%$ for the second source.

\subsection*{Supplementary Discussion 1: Comparing the performance of photon counting and the optimal measurement}
\begin{figure}
    \centering
    \includegraphics[width=0.4\columnwidth]{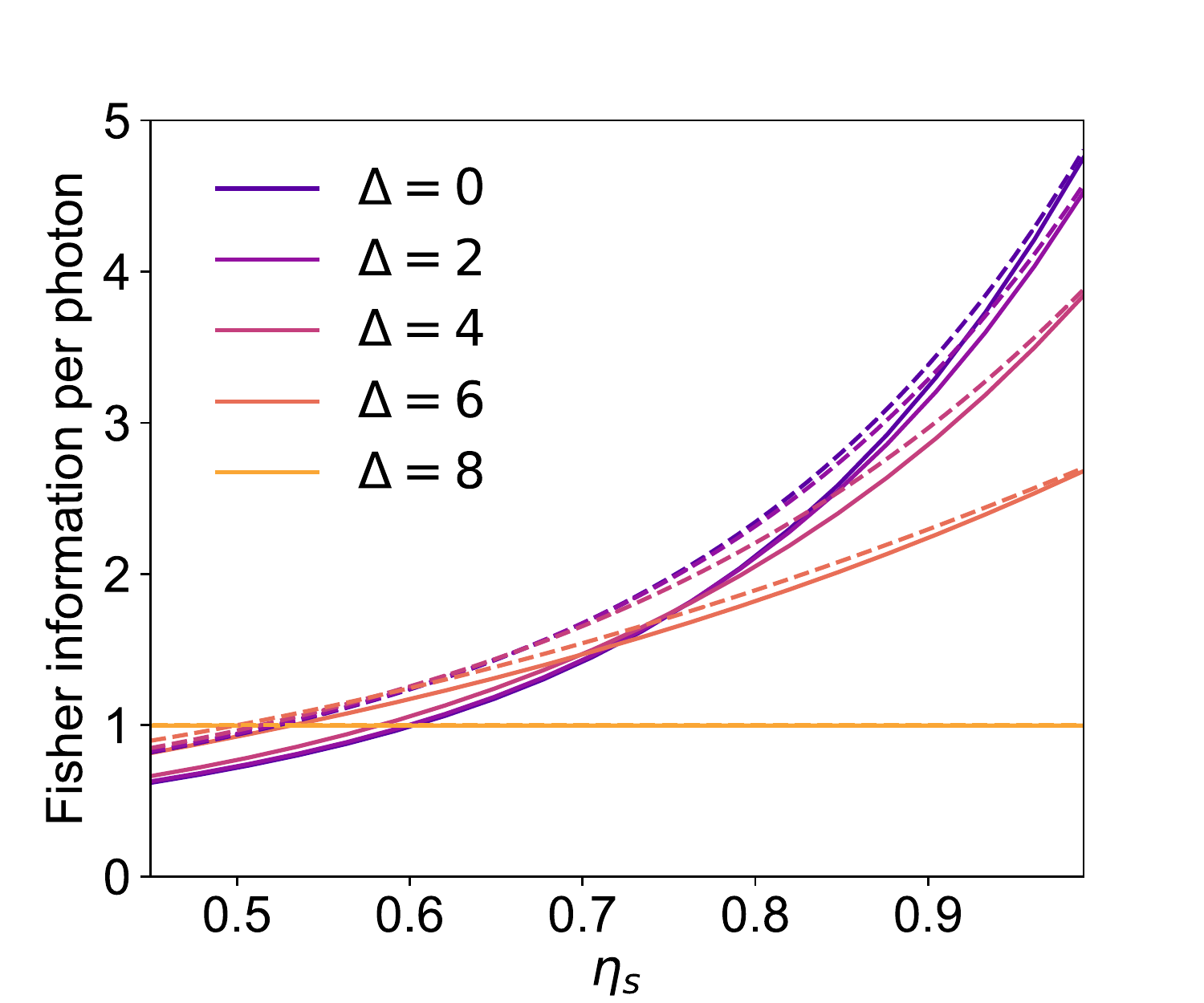}
	\caption{
    Quantum Fisher information per photon, $Q/ \eta_s N$ [dashed lines], and the maximum classical Fisher information per photon, $\mathrm{max}[\tilde{F}(\phi)]$ [continuous lines], for $N=8$ probes with different $\Delta = |h_1-h_2|$.
    We assumed ideal heralding ($\eta_{h_1}=\eta_{h_2}=1$) and balanced loss inside the interferometer ($\eta_{s_1}=\eta_{s_2}=\eta_s$).
    This plot compares the phase sensitivity obtainable with photon counting [continuous lines] to the optimal measurement strategy [dashed lines].
    }
	\label{fig:comparing_qfi_and_cfi}
\end{figure}
Here we compare the performance of our measurement strategy, photon counting, to the optimal measurement strategy.
In Fig.~\ref{fig:comparing_qfi_and_cfi}, we plot the classical Fisher information per photon obtained with photon counting for various $N=8$ probes, assuming ideal heralding ($\eta_{h_1}=\eta_{h_2}=1$) and balanced loss inside the interferometer ($\eta_{s_1}=\eta_{s_2}=\eta_s$).
Since this quantity generally depends on $\phi$ when $\eta_s < 1$, we focus on the region with the largest phase sensitivity, i.e. $\mathrm{max}[\tilde{\mathcal{F}}(\phi)]$ [continuous lines].
In the same plot, we reproduce the quantum Fisher information curves from Fig.~1(b) normalized by the number of detected photons, i.e. $\mathcal{Q}/\eta_s N$ [dashed lines].
For $\eta_s =1$, the quantum and classical Fisher information are equal which means that photon counting is the optimal measurement strategy, as expected~\cite{hofmann2009all}.
For $\eta_s < 1$, photon counting is no longer optimal (except for $\Delta = N$) but still provides quantum-enhanced phase sensitivity for $\eta_s \gtrsim 0.55$.
To the best of our knowledge, the optimal measurement strategy for lossy Holland-Burnett interferometry is not known.
We note that homodyne and weak-field homodyne (i.e. combining a signal with local oscillator then performing photon counting) have been shown to be more loss-tolerant than photon counting using Gaussian probes~\cite{ono2010effects,oh2017practical}.
However, it remains an open question whether these measurement strategies would be advantageous using our non-Gaussian probes in the presence of loss.

\subsection*{Supplementary Discussion 2: Source imperfections}
\begin{figure}
    \centering
	\includegraphics[width=1\columnwidth]{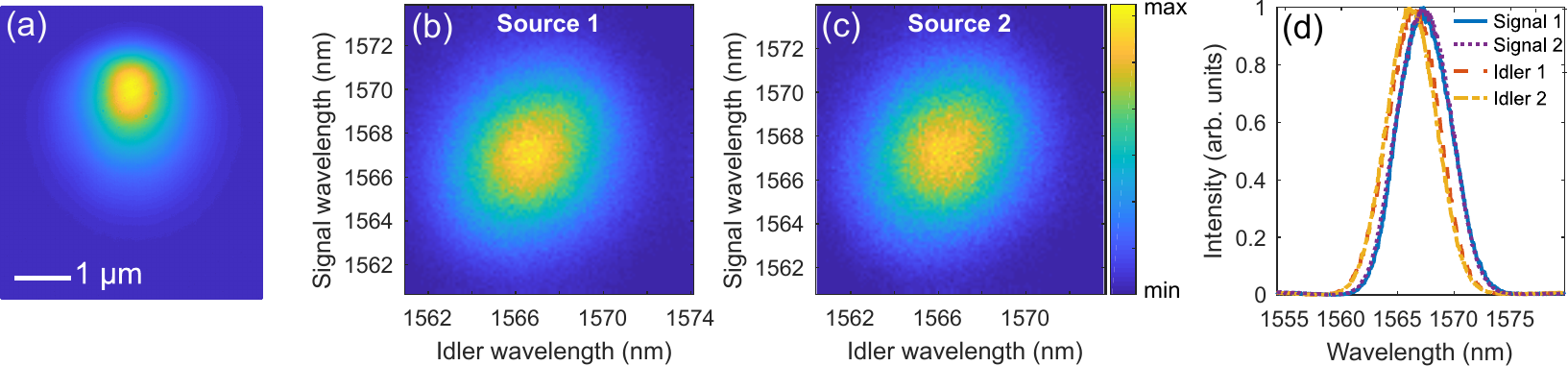}
	\caption{
    Characterization of the sources.
    (a) Typical intensity distribution of the waveguide spatial mode measured at 1550 nm. Its non-Gaussian features limit the fiber coupling efficiency to 70\%. Scale bar shows dimension in the object plane, i.e. at facet of waveguide.
    (b) and (c) Joint spectral intensities of sources 1 and 2, respectively.
    (d) Spectra of the four modes in the experiment.
    }
	\label{fig:source_charac}

\end{figure}

The main imperfections limiting the estimation precision are (i) photon loss and (ii) incomplete interference of the input photons.
The main contribution to photon loss ($\sim$50\% end-to-end, see Supplementary Method 3) are the inefficiences caused by the spatial overlap between the waveguide mode and the fiber mode.
The diffusion process used to produce KTP waveguides leads to non-Gaussian and asymmetric features in the waveguide spatial mode [Fig.~\ref{fig:source_charac}(a)] which limit the fiber coupling efficiency to $\sim$70\%.
It may be possible to improve the spatial mode of the waveguide by optimizing the diffusion process~\cite{padberg2020characterisation} or employing ridge waveguides~\cite{volk2018fabrication}.
The quantum interference visibility ($\sim 75\%$) is mainly limited by spectral mode mismatch between the signal modes as well as the spectral purity of the sources.
We show the joint spectral intensities [Fig.~\ref{fig:source_charac}(b-c)] and marginal spectra [Fig.~\ref{fig:source_charac}(d)] of the sources measured using a time-of-flight spectrometer (resolution $\pm 0.1$nm).
Although the joint spectral intensities appear decorrelated and the signal spectra are well overlapped, we suspect that non-uniform spectral phase (perhaps due to pump chirp or dispersion through optical elements) may have reduced the quantum interference visibility.
Moreover, the spectral purity is degraded by uncorrelated background photons ($\sim 5\%$ of detected photons).
These background photons are generated in a continuum of spectral modes and likely originate from processes where one photon from a down-converted pair is generated in an unguided waveguide mode~\cite{ecksteinthesis}, in which case their contribution could be reduced by minimizing propagation losses inside the waveguide.

\subsection*{Supplementary Discussion 3: Recovering shot-noise limited performance}
\begin{figure}
    \centering
    \includegraphics[width=0.35\columnwidth]{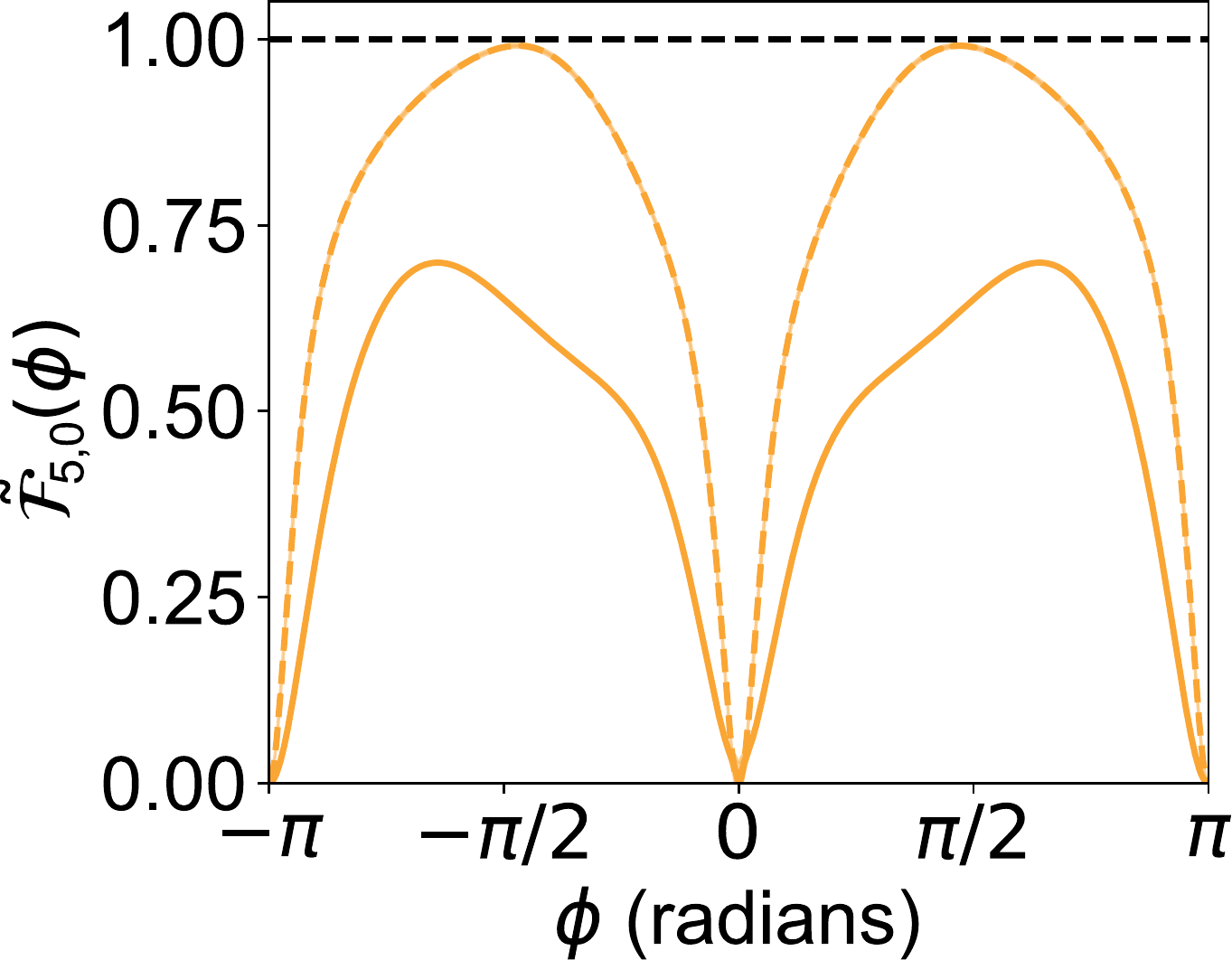}
	\caption{
    $\tilde{\mathcal{F}}_{5,0}(\phi)$ calculated without post-selection.
    Bold yellow line shows result with both sources unblocked, whereas dashed yellow line is result with one of the sources blocked.
    The thicknesses of the yellow lines show $1\sigma$ confidence intervals obtained by fitting 50 simulated experiments.
    Dashed black line is the shot-noise limit.
    }
	\label{fig:SNL_limited}

\end{figure}

Using the $\Delta = N$ probe, all photons injected into the interferometer originate from one source.
As such, imperfections such as spectral purity and mode matching should not affect the performance of the probe.
The $\Delta = N$ probe is prepared by considering trials where e.g. $(h_1,h_2)=(N,0)$.
However, even when $h_2=0$, this second source can still inject unwanted light into the interferometer due to losses in the herald modes.
This generally degrades the performance of the $\Delta=N$ probe.
Here we show that shot-noise limited performance is recovered by blocking one of the sources.
In Fig.~\ref{fig:SNL_limited}, we plot $\tilde{\mathcal{F}}_{5,0}(\phi)$ for the $\Delta = N = 5$ probe calculated without post-selection.
We performed the measurement with a single source blocked and with both sources unblockled.
In the latter case, we find that $\tilde{\mathcal{F}}_{5,0}(\phi)$ reaches $0.991\pm0.001$ at its highest point, demonstrating shot-noise limited performance.
Ideally, $\tilde{\mathcal{F}}_{5,0}(\phi)$ should be flat with $\phi$.
However, experimental imperfections such as imbalanced detector efficiency, detector dark counts ($\sim 1\%$), and imperfect interferometer visibility cause the dips in $\tilde{\mathcal{F}}_{5,0}(0)$ and $\tilde{\mathcal{F}}_{5,0}(\pm \pi)$ where the photons should ideally always exit the interferometer from one port.

\subsection*{Supplementary Discussion 4: Parameters required to surpass shot-noise limit without post-selection}
Here we provide an analysis on estimating the efficiency and quality of the two-mode vacuum sources required to surpass the shot-noise limit without post-selection.
We focus on the $\Delta=6$ [i.e. $(h_1,h_2)=(7,1)$] probe as this is the most loss-tolerant $N=8$ probe that can surpass the shot-noise limit in our scheme.
For simplicity, we assume equal efficiency $\eta$ in all four modes of the experiment ($\eta = \eta_{h_1} = \eta_{h_2} = \eta_{s_1} = \eta_{s_2}$) and equal PDC gain parameters $\lambda$.
There are three main experimental parameters to consider: (i) the efficiency $\eta$, (ii) the distinguishability $\mathcal{M}$ of photons between the top and bottom sources, (iii) the PDC gain $\lambda$.
Given these parameters, we estimate the sensitivity of the probe by calculating the  classical Fisher information $\tilde{\mathcal{F}}_{7,1}(\phi)$ (per detected signal photon).
Since this quantity generally depends on $\phi$, we focus on region in phase with the largest possible sensitivity, i.e. $\max[\tilde{\mathcal{F}}_{7,1}(\phi)]$.

\begin{figure}
    \centering
	\includegraphics[width=0.95\columnwidth]{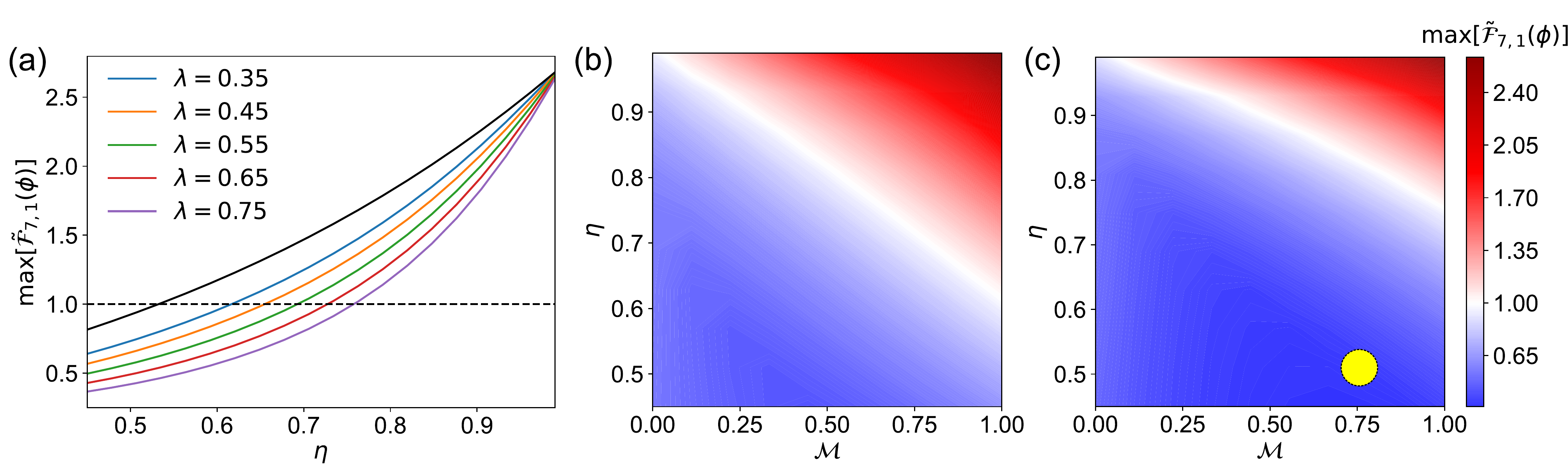}
	\caption{
    (a) Effect of squeezing strength on imperfect heralding. 
    We plot $\mathrm{max}[\tilde{\mathcal{F}}_{7,1}(\phi)]$ as a function of the efficiency $\eta$ (equal in all four modes) for various PDC gain parameters $\lambda$ and assuming $\mathcal{M}=1$.
    Black line shows the perfect heralding case ($\eta_{h_1}=\eta_{h_2}=1$).
    Dashed line shows the shot-noise limit.
    (b) and (c) plots $\mathrm{max}[\tilde{\mathcal{F}}_{7,1}(\phi)]$ as a function of $\eta$ and $\mathcal{M}$ for $\lambda=0.35$ and $\lambda=0.75$, respectively.
    Blue (red) indicates a parameter regime where expected performance is below (above) the shot-noise limit.
    The yellow circle in (c) shows roughly the parameter regime achieved in the experiment.
    }
	\label{fig:proximity_to_snl}
\end{figure}

We begin by focusing on the effect of the PDC gain $\lambda$ and assume $\mathcal{M}=1$ for now. 
As shown in Fig.~\ref{fig:proximity_to_snl}(a), a smaller $\lambda$ provides a larger $\mathrm{max}[\tilde{\mathcal{F}}_{7,1}(\phi)]$.
This is because lowering $\lambda$ increases the photon-number purity of the heralded probe in the presence of loss in the heralding arms, i.e. it reduces the probability that the herald detectors under-counted the true number of photon pairs produced by the sources. 
As a reference, we include the perfect heralding case which is shown by the black line.
While reducing $\lambda$ minimizes the detrimental effects of imperfect heralding, it also drastically decreases the heralding rate.
For example, assuming $\eta=0.5$ and a 100 kHz laser repetition rate, $\lambda=0.75$ would produce a $N=8$ probe roughly once per second whereas $\lambda=0.35$ would produce such a probe only about once per day.

Next we consider the combined effect of imperfect distinguishability and efficiency.
In Fig.~\ref{fig:proximity_to_snl}(b) [(c)], we plot $\mathrm{max}[\tilde{\mathcal{F}}_{7,1}(\phi)]$ as a function of $\eta$ and $\mathcal{M}$ for $\lambda = 0.75$ [$\lambda=0.35$].
The approximate region achieved in our experiment is shown in yellow.
Improvements in both $\eta$ and $\mathcal{M}$ are necessary to unconditionally surpass the shot-noise limit.
As a reference point, a distinguishability of $\mathcal{M}\sim0.85$ was achieved in Ref.~\cite{stobinska2019quantum} using the same type of high-gain PDC sources as used in our experiment.
With such a distinguishability, the efficiency would need to be improved to $\sim 80\%$ [$\sim 70\%$] when using $\lambda=0.75$ [$\lambda=0.35$].

% COMMENT FOR ARXIV

% \bibliographystyle{apsrev4-1}
% \bibliography{refs}

% \end{document}

\end{document}